\documentstyle[epsfig,axodraw,12pt,a4]{article}

\begin{document}
\title{{\bf
QCD Sum Rules: Intercrossed Relations for 
$\Sigma^{0}$ and $ \Lambda$ Magnetic Moments
}}
\author{
{ A.\"{O}zpineci}\thanks{ozpineci@ictp.trieste.it} \\
{\small International Centre for Theoretical Physics, Trieste, Italy} \\
\and
{ S.B.~Yakovlev, and V.S.~Zamiralov} \thanks{zamir@depni.sinp.msu.ru} \\
{\small Institute of Nuclear Physics,} \\
{\small M.V. Lomonosov Moscow State University,} \\
{\small Vorobjovy Gory, Moscow, Russia.}
}

\begin{titlepage}
\maketitle
\thispagestyle{empty}
\begin{abstract}
New relations between QCD Borel sum rules for magnetic
moments of
$\Sigma^{0}$ and $\Lambda$ hyperons are constructed.
It is shown that starting from the sum rule for the
$\Sigma^{0}$ hyperon magnetic moment it is straightforward to obtain
the corresponding sum rule for the $\Lambda$ hyperon magnetic moment
\it{et vice versa}.
\end{abstract}


\newpage


\end{titlepage}
\section{Introduction}

Recently a series of papers were dedicated to study hadron 
properties of the $\Sigma, \Sigma_{c}, \Sigma_{b}$ baryons
as well as of the $\Lambda, \Lambda_{c},\Lambda_{b}$ ones in the framework
of various QCD sum rules \cite{Hwang,Hwang1,Hwang2},
\cite{Altug1,Altug}
(we cite only few examples)
following pioneer works of \cite{Ioffe} and works 
\cite{Pasu,Pasu2,Pasu3}.
Many interesting results were obtained. But full expressions for
mass or magnetic moment sum rules become often too long and tedious to 
achieve and prove. Moreover the $\Lambda$ hyperon
properties is usually treated apart from the other members of the
baryon octet. Other $\Lambda$-like states also are treated apart
from those of the corresponding $\Sigma$-like states.
Is it possible to relate results for $\Lambda$ and $\Sigma$
hyperons among themselves?


We propose here nonlinear intercrossed relations which relate
QCD sum rules for magnetic moments
of $\Sigma$ hyperon with that of
$\Lambda$ one  and {\it vice versa} as well as formulae
relating sum rules for magnetic moment of $\Sigma$ or that of
$\Lambda$ with the corresponding
sum rule for the $\Sigma^{0}\Lambda$ transition magnetic moment. 
Their origin lies in
the relation between isotopic, $U$- and $V$-spin quantities
and is quasi obvious in the framework of the quark model.
These relations 
seem to be useful while obtaining hadron properties
of the $\Lambda$-like baryons from those of the
$\Sigma$-like baryons ( {\it et vice versa}) or checking
expressions for them mutually. 

We begin with a simple example based on the NRQM and then proceed
to the QCD sum rules. 

\section{Relation between magnetic moments of $\Sigma^{0}$ and
$\Lambda$ in the NRQM}

 Let us write magnetic moments
of hyperons $\Sigma^{0}$  and $\Lambda$ of the baryon octet
in the NRQM:
\begin{equation}
\mu(\Sigma^{0}(ud,s))=\frac{2}{3}\mu_{u}+\frac{2}{3}\mu_{d}-
\frac{1}{3}\mu_{s};\quad \mu(\Lambda)=\mu_{s}.
\label{mm8}
\end{equation}
As it is known magnetic moment of any other baryon of the octet but that of
the $\Lambda$ hyperon can be obtained from the expression
for the  $\Sigma^{0}$. E.g.,  magnetic moment of the  $\Sigma^{+}(uu,s)$
hyperon is obtained just by putting $\mu_{u}$ instead of $\mu_{d}$
in Eq.(\ref{mm8}):
$$
\mu(\Sigma^{+})=\frac{4}{3}\mu_{u}-\frac{1}{3}\mu_{s}.
$$
But magnetic moment of the $\Lambda$ hyperon can be also obtained
from that of the $\Sigma^{0}$ one, as well as 
 magnetic moment of the  $\Sigma^{0}$  can be obtained 
from that of the $\Lambda$ one. For that purpose
let us formally perform in Eq.(\ref{mm8}) the exchange $d\leftrightarrow s$ 
to get
\begin{equation}
\mu(\tilde\Sigma^{0}_{d\leftrightarrow s})=
\frac{2}{3}\mu_{u}+\frac{2}{3}\mu_{s}-
\frac{1}{3}\mu_{d};\quad \mu(\tilde\Lambda_{u\leftrightarrow s})=\mu_{d}
\label{sigma}
\end{equation}
and the exchange $u\leftrightarrow s$ to get
\begin{equation}
\mu(\tilde\Sigma^{0}_{u\leftrightarrow s})=
\frac{2}{3}\mu_{d}+\frac{2}{3}\mu_{s}-
\frac{1}{3}\mu_{u};\quad \mu(\tilde\Lambda_{u\leftrightarrow s})=\mu_{u}.
\label{lambda}
\end{equation}
The following relations are valid:
\begin{eqnarray}
2(\mu(\tilde\Sigma^{0}_{d\leftrightarrow s})+
\mu(\tilde\Sigma^{0}_{u\leftrightarrow s}))-
\mu(\Sigma^{0})=3\mu(\Lambda);\\
2(\mu(\tilde\Lambda_{d\leftrightarrow s})+
\mu(\tilde\Lambda_{u\leftrightarrow s}))-
\mu(\Lambda)=3\mu(\Sigma^{0})\nonumber.
\end{eqnarray}
Also
\begin{eqnarray}
\mu(\tilde\Sigma^{0}_{d\leftrightarrow s})-
\mu(\tilde\Sigma^{0}_{u\leftrightarrow s})=
\sqrt{3}\mu(\Sigma^{0}\Lambda);\\
\mu(\tilde\Lambda_{d\leftrightarrow s})-
\mu(\tilde\Lambda_{u\leftrightarrow s})=
-\sqrt{3}\mu(\Sigma^{0}\Lambda)\nonumber.
\end{eqnarray}
The origin of these relations lies in the structure of
baryon wave functions in the NRQM with isospin $I=1,0$
and $I_{3}=0$:
$$
2\sqrt{3}|\Sigma^{0}(ud,s)>_{\uparrow}=
|2u_{\uparrow}d_{\uparrow}s_{\downarrow}+
2d_{\uparrow}u_{\uparrow}s_{\downarrow}-
u_{\uparrow}s_{\uparrow}d_{\downarrow}-
s_{\uparrow}u_{\uparrow}d_{\downarrow}-
d_{\uparrow}s_{\uparrow}u_{\downarrow}-
s_{\uparrow}d_{\uparrow}u_{\downarrow}>,
$$
$$
2|\Lambda>_{\uparrow}=
|d_{\uparrow}s_{\uparrow}u_{\downarrow}+
s_{\uparrow}d_{\uparrow}u_{\downarrow}-
u_{\uparrow}s_{\uparrow}d_{\downarrow}-
s_{\uparrow}u_{\uparrow}d_{\downarrow}>,
$$
where $q_{\uparrow}$ ($q_{\downarrow}$) means wave function of
the quark $q$ (here $q=u,d,s$) with the helicity +1/2 (-1/2).
With the exchanges $d\leftrightarrow s$ and $u\leftrightarrow s$ 
one arrives at the corresponding $U$-spin and $V$-spin quantities, so
$U=1,0$ and $U_{3}=0$ baryon wave functions are
$$
-2|\tilde\Sigma^{0}_{d\leftrightarrow s}(us,d)>=
|\Sigma^{0}(ud,s)>+\sqrt{3}|\Lambda>,
$$
$$
-2|\tilde\Lambda_{d\leftrightarrow s}>=
-\sqrt{3}|\Sigma^{0}(ud,s)>+|\Lambda>,
$$
while $V=1,V_{3}=0$ and $V=0$ baryon wave functions are
$$
-2\tilde\Sigma^{0}_{u\leftrightarrow s}(ds,u)=
|\Sigma^{0}(ud,s)>-\sqrt{3}|\Lambda>,
$$
$$
2|\tilde\Lambda_{u\leftrightarrow s}>=\sqrt{3}|\Sigma^{0}(ud,s)>+|\Lambda>.
$$
It is easy to show that relations given by 
Eqs.(\ref{sigma},\ref{lambda})
immediately follow.

\section{Relation between QCD correlators for 
$\Sigma^{0}$ and $\Lambda$ }

Now we demonstrate how similar considerations work for QCD sum rules
on the example of QCD Borel sum rules for magnetic moments.

The starting point would be two-point Green's function
for hyperons $\Sigma^{0}$ and $\Lambda$ of the baryon octet:
\begin{equation}
\Pi^{\Sigma^{0},\Lambda}=i\int d^{4}x e^{ipx}
<0|T\{ {\eta^{\Sigma^{0},\Lambda}(x),\eta^{\Sigma^{0},\Lambda}(0)}\}|0>,
\label{two}
\end{equation}
where isovector (with $I_{3}=0$) and isocalar field operators 
could be chosen as \cite{Altug1}
\begin{eqnarray}
\eta^{\Sigma^{0}}=\frac{1}{2}\epsilon_{abc}
[\large(u^{aT}Cs^{b}\large)\gamma_{5}d^{c}-
\large(d^{aT}Cs^{b}\large)\gamma_{5}u^{c}-
\large(u^{aT}C\gamma_{5}s^{b}\large)d^{c}+
\large(d^{aT}C\gamma_{5}s^{b}\large)u^{c}],
\nonumber\\
\eta^{\Lambda}=\frac{1}{2\sqrt{3}}\epsilon_{abc}
[-2\large(u^{aT}Cd^{b}\large)\gamma_{5}s^{c}+
\large(u^{aT}Cs^{b}\large)\gamma_{5}d^{c}+
\large(d^{aT}Cs^{b}\large)\gamma_{5}u^{c}+
\nonumber\\
2\large(u^{aT}C\gamma_{5}d^{b}\large)s^{c}-
\large(u^{aT}C\gamma_{5}s^{b}\large)d^{c}-
\large(d^{aT}C\gamma_{5}s^{b}\large)u^{c}],\hspace{10mm}
\label{eta}
\end{eqnarray}
where $a,b,c$ are the color indices and $
u,d,s$ are quark wave functions, $C$ is charge conjugation matrix,

We show now that one can operate with $\Sigma$ hyperon
and obtain the results
for the $\Lambda$ hyperon. The reasoning would be valid also for 
charm and beauty $\Sigma$-like and  $\Lambda$-like baryons.

In order to arrive at the desired relations we write not
only isospin quantities but also  $U$-spin and  $V$-spin ones.

Let us introduce $U$-vector 
(with $U_{3}=0$) and $U$-scalar field operators just
formally changing $(d\leftrightarrow s)$ in the Eq.(\ref{eta}):
\begin{eqnarray}
\tilde\eta^{\Sigma^{0}(d\leftrightarrow s)}
=\frac{1}{2}\epsilon_{abc}
[\large(u^{aT}C\cdot 1\cdot d^{b}\large)\gamma_{5}s^{c}-
\large(s^{aT}Cd^{b}\large)\cdot\gamma_{5}\cdot u^{c}-
(1 \leftrightarrow \gamma_{5})]
\nonumber\\
\tilde\eta^{\Lambda(d\leftrightarrow s)}=
\frac{1}{2\sqrt{3}}\epsilon_{abc}
[\large(-2\large(u^{aT}C\cdot 1\cdot s^{b}\large)\gamma_{5}d^{c}+
\large(u^{aT}C\cdot 1\cdot d^{b}\large)\cdot\gamma_{5}\cdot s^{c}+
\nonumber\\
\large(s^{aT}C\cdot 1\cdot d^{b}\large)\cdot\gamma_{5}\cdot u^{c}\large)-
(1 \leftrightarrow \gamma_{5})].
\label{tildeta}
\end{eqnarray}
Similarly we introduce $V$-vector 
(with $V_{3}=0$) and $V$-scalar field operators just
changing $(u\leftrightarrow s)$ in the Eq.(\ref{eta}):
\begin{eqnarray}
\tilde\eta^{\Sigma^{0}(u\leftrightarrow s)}
=\frac{1}{2}\epsilon_{abc}
[\large(s^{aT}C\cdot 1\cdot u^{b}\large)\gamma_{5}d^{c}-
\large(d^{aT}Cu^{b}\large)\cdot\gamma_{5}\cdot s^{c}-
(1 \leftrightarrow \gamma_{5})]
\nonumber\\
\tilde\eta^{\Lambda(u\leftrightarrow s)}=
\frac{1}{2\sqrt{3}}\epsilon_{abc}
[\large(-2\large(u^{aT}C\cdot 1\cdot s^{b}\large)\gamma_{5}u^{c}+
\large(u^{aT}C\cdot 1\cdot u^{b}\large)\cdot\gamma_{5}\cdot s^{c}+
\nonumber\\
\large(d^{aT}C\cdot 1\cdot u^{b}\large)\cdot\gamma_{5}\cdot s^{c}\large)-
(1 \leftrightarrow \gamma_{5})].
\label{tildet}
\end{eqnarray}
Field operators of the Eq.(\ref{eta}) and Eq.(\ref{tildeta})
can be related through
\begin{eqnarray}
-2\tilde\eta^{\Lambda(d\leftrightarrow s)}=
\sqrt{3}\eta^{\Sigma^{0}}+\eta^{\Lambda},
\nonumber\\
2\tilde\eta^{\Sigma^{0}(d\leftrightarrow s)}=
\eta^{\Sigma^{0}}-\sqrt{3}\eta^{\Lambda},
\nonumber\\
2\tilde\eta^{\Lambda(u\leftrightarrow s)}=
\sqrt{3}\eta^{\Sigma^{0}}\eta^{\Lambda},
\nonumber\\
2\tilde\eta^{\Sigma^{0}(u\leftrightarrow s)}=
\eta^{\Sigma^{0}}+\sqrt{3}\eta^{\Lambda}.
\label{etarel}
\end{eqnarray}

Upon using Eqs.(\ref{eta}-\ref{etarel})
two-point functions of the Eq.(\ref{two}) for 
hyperons $\Sigma^{0}$ and $\Lambda$ of the baryon octet
can be related as
\begin{equation}
2[\tilde\Pi^{\Sigma^{0}(d\leftrightarrow s)}+
\tilde\Pi^{\Sigma^{0}(u\leftrightarrow s)}]-
\Pi^{\Sigma^{0}}=3\Pi^{\Lambda},
\label{crossSi}
\end{equation}

\begin{equation}
2[\tilde\Pi^{\Lambda(d\leftrightarrow s)}+
\Pi^{\tilde\Lambda(u\leftrightarrow s)}]-
\Pi^{\Lambda}=3\Pi^{\Sigma^{0}}.
\label{crossLa}
\end{equation}

These are essentially nonlinear relations.
 
It is seen that starting calculations, e.g., from 
$\Sigma$-like quantities one should arrive 
at the corresponding quantities
for $\Lambda$-like baryons {\it et vice versa}.

Moreover one can obtain QCD Borel sum rule for the 
$\Sigma^{0}-\Lambda$ transition magnetic moment using
the relations
\begin{eqnarray} 
2 \left[\tilde\Pi^{\Sigma^{0}(u\leftrightarrow s)}-
\tilde\Pi^{\Sigma^{0}(d\leftrightarrow s)} \right]
=\sqrt{3} \left( \Pi^{\Lambda\Sigma^{0}} + \Pi^{\Sigma^{0}\Lambda} \right),
\label{crossSiLa} \\
%
2 \left[\tilde\Pi^{\Lambda(u\leftrightarrow s)}-
\Pi^{\tilde\Lambda(d\leftrightarrow s)} \right]
=-\sqrt{3}\left( \Pi^{\Lambda\Sigma^{0}} + \Pi^{\Sigma^{0}\Lambda} \right).
\label{crossLaSi}
\end{eqnarray} 

\section{Intercrossed relations for the QCD 
magnetic moment sum rules}

In order to demonstrate clearly how it works we preferred not to use
QCD sum rules elaborated by one of us with coauthors \cite{Altug1},
\cite{Altug} though they perfectly satisfy
the relations (\ref{crossSi}-\ref{crossLaSi}), but instead
to repeat calculations of the first of the QCD Borel sum rules 
for   magnetic moments following \cite{Pasu} conserving 
non-degenerated quantities for $u$ and $d$ quarks. 

We would need quantities \cite{Ioffe,Pasu}
\begin{eqnarray}
a_{q}=-(2\pi)^{2}<\bar q q>,\quad
b=<g_{c}G^{2}>,
\nonumber\\
a_{q}m_{0(q)}^{2}=(2\pi)^{2}<g_{c}\bar q \sigma\cdot G q>,\quad q=u,d,s.
\nonumber\\
<\bar q \sigma_{\mu\nu}q>_{F}=e_{q}\chi<\bar q q>F_{\mu\nu},
\label{vev}
\end{eqnarray}
while the susceptibilities $\kappa$ and $\xi$ are defined through
$$
<\bar q g_{s}G_{\mu\nu}q>_{F}=e_{q}\kappa<\bar q q>F_{\mu\nu},
$$
$$
\epsilon_{\alpha\beta\mu\nu}
<\bar q g_{s}G^{\mu\nu}\gamma_{5}q>_{F}=
ie_{q}\xi<\bar q q>F_{\alpha\beta}.
$$
Also we would define a factor used to subtract the continuum
contribution \cite{Ioffe}:
$$
E_{n}(x)=1-e^{-x}(1+x+...+x^{n}/n!),\quad x=W_{B}^{2}/M^{2}, \quad 
B=\Sigma^{0},\Lambda,
$$
although here we do not use it.
 
We shall
construct sum rule for the magnetic moment of the
$\Sigma^{0}$ hyperon in the form
\begin{eqnarray}
SR (\Sigma^{0})=\sum_{i=1}^{10}\Sigma^{0(i)}=
\beta^{2}_{\Sigma^{0}}(\mu(\Sigma^{0})+AM^{2})
e^{-M^{2}_{\Sigma^{0}}/M^{2}}+e.s.c.,
\label{hwangSi}
\end{eqnarray}
and then (re)derive term by term the corresponding quantities for the
$\Lambda$ hyperon upon using Eq.(\ref{crossSi}) and
$\Sigma^{0}-\Lambda$ transition upon using Eq.(\ref{crossSiLa}).
Here $\beta_{\Sigma^{0}}$ is a coupling strength of the
$\Sigma^{0}$ current to the ground state of this hyperon,
while $A$ is a constant arising from the non-diagonal
transitions and $e.s.c.$ means 'excited state contributions.

We would proceed pass by pass to show also that every group
of diagrams yields expressions which can be treated 
through  Eqs.(\ref{crossSi},\ref{crossSiLa}) in an autonomic way.

The 1st term $\Sigma^{0(1)}$ comes from the first two
diagrams of the Fig.1.
\vskip 5mm
\begin{center}
\begin{picture}(400,100)(0,0)
\CArc(20,40)(20,90,270)
\CArc(60,40)(20,270,90)
\Line(20,60)(60,60)
\Line(20,20)(60,20)
\Line(0,40)(80,40)
\Photon(40,60)(40,90)3 3
\Text(25,70)[c]{u}
\Text(30,50)[c]{d}
\Text(50,10)[c]{s}
\CArc(120,40)(20,90,270)
\CArc(160,40)(20,270,90)
\Line(120,60)(160,60)
\Line(120,20)(160,20)
\Line(100,40)(180,40)
\Photon(140,60)(140,90)3 3
\Text(125,70)[c]{d}
\Text(130,50)[c]{u}
\Text(150,10)[c]{s}
\CArc(220,40)(20,90,270)
\CArc(260,40)(20,270,90)
\Line(220,60)(260,60)
\Line(220,20)(260,20)
\Line(200,40)(280,40)
\Photon(240,20)(240,0)3 3
\Text(225,70)[c]{u}
\Text(230,50)[c]{d}
\Text(250,10)[c]{s}
\Text(190,0)[c]{Fig.1}
\end{picture}
\end{center}
$$
2\Sigma^{0(1)}=\frac{M^{6}}{4L^{4/9}}
2(e_{u}+e_{d}),
$$
where from the auxiliary quantities
$\tilde\Sigma^{0(1)}_{sd}$ and $\tilde\Sigma^{0(1)}_{su}$ 
would be
$$
2\tilde\Sigma^{0(1)}_{sd}=\frac{M^{6}}{4L^{4/9}}
2(e_{u}+e_{s}),
\quad 2\tilde\Sigma^{0(1)}_{su}=\frac{M^{6}}{4L^{4/9}}
2(e_{d}+e_{s}).
$$
Applying Eq.(\ref{crossSi}) we arrive at the contribution
of the $\Lambda$ hyperon $\Lambda^{(1)}$ given already by all
three diagrams of the Fig.1:
$$
\Lambda^{(1)}=
\frac{M^{6}}{12L^{4/9}}(e_{u}+e_{d}+4e_{s})
$$
in agreement with the 1st term of the Eq.(29) 
in \cite{Pasu}.
The difference between these auxiliary quantities just gives
up to a factor $\sqrt{3}$ the $\Sigma^{0}\Lambda$ 
transition magnetic moment
in accordance with the relation (\ref{crossSiLa}):
$$
\sqrt{3}(\Sigma^{0}\Lambda)^{(1)}=\tilde\Sigma^{0(1)}_{su}-
\tilde\Sigma^{0(1)}_{sd}=
\frac{M^{6}}{4L^{4/9}}
(e_{u}-e_{d}),
$$
which agrees with the corresponding term in \cite{Hwang2}.

The 2nd term $\Sigma^{0(2)}$ comes from the
diagrams of the Fig.2. 
\vskip 5mm
\begin{center}
\begin{picture}(400,100)(0,0)
\CArc(20,40)(20,90,270)
\CArc(60,40)(20,270,90)
\Line(40,60)(60,60)
\Line(20,20)(60,20)
\Line(0,40)(30,40)
\Line(50,40)(80,40)
\Photon(60,60)(60,90)3 3
\Text(20,70)[c]{u}
\Text(20,50)[c]{d}
\Text(20,30)[c]{s}
\CArc(120,40)(20,90,270)
\CArc(160,40)(20,270,90)
\Line(140,60)(160,60)
\Line(120,20)(160,20)
\Line(100,40)(130,40)
\Line(150,40)(180,40)
\Photon(160,60)(160,90)3 3
\Text(120,70)[c]{d}
\Text(120,50)[c]{u}
\Text(120,30)[c]{s}
\CArc(220,40)(20,90,270)
\CArc(260,40)(20,270,90)
\Line(220,60)(230,60)
\Line(250,60)(260,60)
\Line(220,20)(260,20)
\Line(200,40)(230,40)
\Line(250,40)(280,40)
\Photon(240,20)(240,0)3 3
\Text(220,70)[c]{d}
\Text(220,50)[c]{u}
\Text(220,30)[c]{s}
\Text(190,0)[c]{Fig.2}
\end{picture}
\end{center}
$$
2\Sigma^{0(2)}=
-\frac{L^{4/9}}{18M^{2}}a_{u}a_{d}
2(e_{u}+e_{d}+3e_{s});
$$
$$
2\tilde\Sigma^{0(2)}_{sd}=-\frac{L^{4/9}}{18M^{2}}a_{u}a_{s}
2(e_{u}+e_{s}+3e_{d});
$$
$$
2\tilde\Sigma^{0(2)}_{su}=-\frac{L^{4/9}}{18M^{2}}a_{d}a_{s}
2(e_{d}+e_{s}+3e_{u}).
$$
Upon applying Eq.(\ref{crossSi}) we obtain
$$
\Lambda^{(2)}=\frac{1}{3}[2\tilde\Sigma^{0(2)}_{sd}+
2\tilde\Sigma^{0(2)}_{su}-\Sigma^{0(2)}]=
$$
$$
-\frac{L^{4/9}}{54M^{2}}\{2[(e_{u}a_{u}+e_{d}a_{d})+
3(e_{u}a_{d}+e_{d}a_{u})+e_{s}(a_{u}+a_{d})]-
$$
$$
a_{u}a_{d}(e_{u}+e_{d}+3e_{s})\}.
$$
Taking $a_{u}=a_{d}=a$, $a_{s}/a=f+1$, we get
$$
\Lambda^{(2)}=-\frac{L^{4/9}}{108M^{2}}
[2(7e_{u}+7e_{d}+e_{s})+8f(2e_{u}+2e_{d}+e_{s})]a^{2}
$$
in agreement with the Eq.(29) in \cite{Pasu}.

The difference between these auxiliary quantities yields:
$$
\sqrt{3}(\Sigma^{0}\Lambda)^{(2)}=
\frac{L^{4/9}}{18M^{2}}a_{s}[(a_{u}e_{u}-a_{d}e_{d})+
3(e_{d}a_{u}-e_{u}a_{d})+e_{s}(a_{u}-a_{d})]
$$
$$
\rightarrow (\quad a_{u}=a_{d}=a) 
-\frac{L^{4/9}}{9M^{2}}a_{s}a(e_{u}-e_{d}).
$$
in agreement with the corresponding term in \cite{Hwang2}.

The 3nd term $\Sigma^{0(3)}$ comes from the convergent part of
the diagrams of the Fig.3-5.
\vskip 5mm
\begin{center}
\begin{picture}(400,100)(0,0)
\CArc(20,40)(20,90,270)
\CArc(60,40)(20,270,90)
\Line(20,60)(60,60)
\Line(20,20)(60,20)
\Line(0,40)(80,40)
\Photon(40,60)(40,90)3 3
\DashLine(40,40)(60,70)2
\DashLine(50,20)(70,60)2
\Text(40,0)[c]{Fig.3}
\CArc(120,40)(20,90,270)
\CArc(160,40)(20,270,90)
\Line(120,60)(160,60)
\Line(120,20)(160,20)
\Line(100,40)(180,40)
\Photon(140,60)(140,90)3 3
\DashLine(150,60)(170,80)2
\DashLine(150,40)(170,60)2
\Text(140,0)[c]{Fig.4}
\CArc(220,40)(20,90,270)
\CArc(260,40)(20,270,90)
\Line(220,60)(260,60)
\Line(220,20)(260,20)
\Line(200,40)(280,40)
\Photon(240,60)(240,90)3 3
\DashLine(230,60)(230,80)2
\DashLine(250,60)(250,80)2
\Text(240,0)[c]{Fig.5}
\end{picture}
\end{center}
$$
2\Sigma^{0(3)}=\frac{M^{2}b}{24L^{4/9}}
(2e_{u}+2e_{d}+e_{s}).
$$
It is straightforward to obtain from Eq.(\ref{crossSi})
$$
\Lambda^{(3)}=
\frac{M^{2}b}{144L^{4/9}}(4e_{u}+4e_{d}+7e_{s})
$$
again in agreement with the Eq.(29) in \cite{Pasu}.
The difference between these auxiliary quantities gives:
$$
\sqrt{3}(\Sigma^{0}\Lambda)^{(3)}=
\frac{M^{2}b}{24L^{4/9}}(e_{u}-e_{d}).
$$
The 4th term comes from the divergent part of
the diagrams of the Fig.4, regularized by a cutoff $\Lambda$:
$$
2\Sigma^{0(4)}=\frac{-M^{2}b}{144L^{4/9}}
[ln(\frac{M^{2}}{\Lambda^{2}})
-1-\gamma_{EM}]
2(e_{u}+e_{d}+2e_{s}).
$$
It is straightforward to obtain from Eq.(\ref{crossSi})
$$
\Lambda^{(4)}=
\frac{-M^{2}b}{192L^{4/9}}[ln(\frac{M^{2}}{\Lambda^{2}})
-1-\gamma_{EM}]
(5e_{u}+5e_{d}+2e_{s})
$$
again in full agreement with the 4th term of the
Eq.(29) in \cite{Pasu}.

The difference $\tilde\Sigma^{0(4)}_{su}-\tilde\Sigma^{0(4)}_{sd}$
yields:
$$
\sqrt{3}(\Sigma^{0}\Lambda)^{(4)}=
-\frac{M^{2}b}{144L^{4/9}}
[ln(\frac{M^{2}}{\Lambda^{2}})
-1-\gamma_{EM}]
(e_{u}-e_{d}).
$$
The 5th term comes from the divergent part of
the diagrams of the Fig.5, regularized by a cutoff $\Lambda$:
$$
2\Sigma^{0(5)}=-\frac{M^{2}b}{36L^{4/9}}
[ln(\frac{M^{2}}{\Lambda^{2}})
-\gamma_{EM}-\frac{M^{2}}{2\Lambda^{2}}]
2(e_{u}+e_{d}).
$$
We readily obtain from Eq.(\ref{crossSi})
$$
\Lambda^{(5)}=
-\frac{M^{2}b}{108L^{4/9}}[ln(\frac{M^{2}}{\Lambda^{2}})
-\gamma_{EM}-\frac{M^{2}}{2\Lambda^{2}}]
(e_{u}+e_{d}+4e_{s})
$$
in agreement with the result of \cite{Pasu}.

The difference $\tilde\Sigma^{0(5)}_{su}-\tilde\Sigma^{0(5)}_{sd}$
yields:
$$
\sqrt{3}(\Sigma^{0}\Lambda)^{(5)}=
-\frac{M^{2}b}{36L^{4/9}}
[ln(\frac{M^{2}}{\Lambda^{2}})
-\gamma_{EM}-\frac{M^{2}}{2\Lambda^{2}}]
(e_{u}-e_{d}).
$$

The 6th term $\Sigma^{0(6)}$ involves susceptibilities 
$\chi$, $\kappa$, $\xi$
as it is seen from the diagrams of the Fig.6 and is given by
\vskip 5mm
\begin{center}
\begin{picture}(400,100)(0,0)
\CArc(20,40)(20,90,270)
\CArc(60,40)(20,270,90)
\Line(20,60)(30,60)
\Line(50,60)(60,60)
\Line(20,20)(60,20)
\Line(0,40)(30,40)
\Line(50,40)(80,40)
\DashLine(40,20)(40,40)2
\Photon(40,60)(40,90)3 3
\Text(25,70)[c]{u}
\Text(30,50)[c]{d}
\Text(50,10)[c]{s}
\CArc(120,40)(20,90,270)
\CArc(160,40)(20,270,90)
\Line(120,60)(130,60)
\Line(150,60)(160,60)
\Line(120,20)(160,20)
\Line(100,40)(130,40)
\Line(150,40)(180,40)
\Photon(140,60)(140,90)3 3
\DashCArc(130,40)(10,180,0)2
\Text(125,70)[c]{u}
\Text(130,50)[c]{d}
\Text(150,10)[c]{s}
\CArc(220,40)(20,90,270)
\CArc(260,40)(20,270,90)
\Line(220,60)(230,60)
\Line(250,60)(260,60)
\Line(220,20)(260,20)
\Line(200,40)(230,40)
\Line(250,40)(280,40)
\DashLine(240,20)(240,40)2
\Photon(240,60)(240,90)3 3
\Text(225,70)[c]{d}
\Text(230,50)[c]{u}
\Text(250,10)[c]{s}
\CArc(320,40)(20,90,270)
\CArc(360,40)(20,270,90)
\Line(320,60)(330,60)
\Line(350,60)(360,60)
\Line(320,20)(360,20)
\Line(300,40)(330,40)
\Line(350,40)(380,40)
\Photon(340,60)(340,90)3 3
\DashCArc(330,40)(10,180,0)2
\Text(325,70)[c]{d}
\Text(330,50)[c]{u}
\Text(350,10)[c]{s}
\Text(190,0)[c]{Fig.6}
\end{picture}
\end{center}
$$
2\Sigma^{0(6)}=-\frac{1}{3 L^{4/27}}[
(M^{2}-\frac{m_{(d)0}^{2}}{8L^{4/9}})
e_{u}(\chi_{u}a_{u})a_{d}+
(M^{2}-\frac{m_{(u)0}^{2}}{8L^{4/9}})
e_{d}(\chi_{d}a_{d})a_{u})]+
$$
$$
\frac{L^{4/9}}{18}[
e_{u}((2\kappa_{u}-\xi_{u})a_{u}a_{d}+
e_{d}((2\kappa_{d}-\xi_{d})a_{d}a_{u})],
$$
while the auxiliary quantities are:
$$
2\tilde\Sigma^{0(6)}_{sd}=-\frac{1}{3 L^{4/27}}[
(M^{2}-\frac{m_{(s)0}^{2}}{8L^{4/9}})
e_{u}(\chi_{u}a_{u})a_{s}+
(M^{2}-\frac{m_{(u)0}^{2}}{8L^{4/9}})
e_{s}(\chi_{s}a_{s})a_{u}]+
$$
$$
\frac{L^{4/9}}{18}[
e_{u}((2\kappa_{u}-\xi_{u})a_{u}a_{s}+
e_{s}((2\kappa_{s}-\xi_{s})a_{s}a_{u})];
$$
$$
2\tilde\Sigma^{0(6)}_{su}=-\frac{1}{3 L^{4/27}}[
(M^{2}-\frac{m_{(s)0}^{2}}{8L^{4/9}})
e_{d}(\chi_{d}a_{d})a_{s}+
(M^{2}-\frac{m_{(d)0}^{2}}{8L^{4/9}})
e_{s}(\chi_{s}a_{s})a_{d}]+
$$
$$
\frac{L^{4/9}}{18}[
e_{d}((2\kappa_{d}-\xi_{d})a_{d}a_{s}+
e_{s}((2\kappa_{s}-\xi_{s})a_{s}a_{d})].
$$
Upon using Eq.(\ref{crossSi}) we get:
$$
\Lambda^{(6)}=
-\frac{1}{18 L^{4/27}}\{
(M^{2}-\frac{m_{(d)0}^{2}}{8L^{4/9}})[2e_{s}(\chi_{s}a_{s})
-e_{u}(\chi_{u}a_{u})]a_{d}+
$$
$$
(M^{2}-\frac{m_{(u)0}^{2}}{8L^{4/9}})[2e_{s}(\chi_{s}a_{s})
-e_{d}(\chi_{d}a_{d})]a_{u}
+(M^{2}-\frac{m_{(s)0}^{2}}{8L^{4/9}})[2e_{u}(\chi_{u}a_{u})
+2e_{d}(\chi_{d}a_{d})]a_{s}\}+
$$
$$
\frac{L^{4/9}}{108}\{e_{u}(2\kappa_{u}-\xi_{u} )a_{u}(2a_{s}-a_{d})
+e_{d}(2\kappa_{d}-\xi_{d} )a_{d}(2a_{s}-a_{u})+
2e_{s}(2\kappa_{s}-\xi_{s} )a_{s}(a_{u}+a_{d})\},
$$
which with $a_{u}=a_{d}=a$, $\chi_{u}=\chi_{d}=\chi$,
$\kappa_{u}=\kappa_{d}=\kappa $, $\xi_{u}=\xi_{d}=\xi$,
and
$a_{s}/a=f+1$, $(\chi_{s}a_{s})/(\chi a)=\phi $, similar for
$\xi, \kappa$,  goes to
$$
\rightarrow [-\frac{\chi a^{2}}{18 L^{4/27}}
(M^{2}-\frac{m_{(s)0}^{2}}{8L^{4/9}})+\frac{L^{4/9}}{108}
(2\kappa-\xi )a^{2}][(e_{u}+e_{d})(1+2f)+4e_{s}\phi]
$$
in accord with \cite{Pasu}.
The difference $\tilde\Sigma^{0(6)}_{su}-\tilde\Sigma^{0(6)}_{sd}$
yields:
$$
\sqrt{3}(\Sigma^{0}\Lambda)^{(6)}\rightarrow -(e_{u}-e_{d})
[\frac{-\chi }{6 L^{4/27}}
(M^{2}-\frac{m_{(s)0}^{2}}{8L^{4/9}})+\frac{L^{4/9}}{36}
(2\kappa-\xi ) ]a a_{s}
$$
which agrees with the result of \cite{Hwang2}. 

Now we perform calculations of the 7th term given by the diagrams 
of the Fig.7:
\vskip 5mm
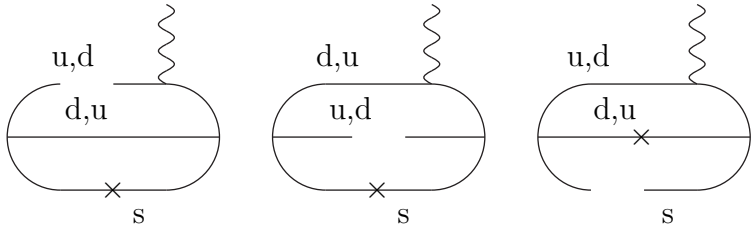
\begin{center}
\begin{picture}(400,100)(0,0)
\CArc(20,40)(20,90,270)
\CArc(60,40)(20,270,90)
\Line(40,60)(60,60)
\Line(20,20)(60,20)
\Line(0,40)(80,40)
\Photon(60,60)(60,90)3 3
\Text(40,20)[c]{$\times$}
\Text(25,70)[c]{u,d}
\Text(30,50)[c]{d,u}
\Text(50,10)[c]{s}
\CArc(120,40)(20,90,270)
\CArc(160,40)(20,270,90)
\Line(120,60)(160,60)
\Line(120,20)(160,20)
\Line(100,40)(130,40)
\Line(150,40)(180,40)
\Photon(160,60)(160,90)3 3
\Text(140,20)[c]{$\times$}
\Text(125,70)[c]{d,u}
\Text(130,50)[c]{u,d}
\Text(150,10)[c]{s}
\CArc(220,40)(20,90,270)
\CArc(260,40)(20,270,90)
\Line(220,60)(260,60)
\Line(240,20)(260,20)
\Line(200,40)(280,40)
\Photon(260,60)(260,90)3 3
\Text(240,40)[c]{$\times$}
\Text(220,70)[c]{u,d}
\Text(230,50)[c]{d,u}
\Text(250,10)[c]{s}
\end{picture}
\end{center}
\begin{center}
\begin{picture}(400,100)(0,0)
\CArc(20,40)(20,90,270)
\CArc(60,40)(20,270,90)
\Line(40,60)(60,60)
\Line(20,20)(60,20)
\Line(0,40)(80,40)
\Photon(60,60)(60,90)3 3
\Text(40,40)[c]{$\times$}
\Text(25,70)[c]{u,d}
\Text(30,50)[c]{d,u}
\Text(50,10)[c]{s}
\CArc(220,40)(20,90,270)
\CArc(260,40)(20,270,90)
\Line(220,60)(260,60)
\Line(240,20)(260,20)
\Line(200,40)(280,40)
\Photon(260,20)(260,0)3 3
\Text(240,60)[c]{$\times$}
\Text(220,70)[c]{u,d}
\Text(230,50)[c]{d,u}
\Text(250,10)[c]{s}
\Text(190,0)[c]{Fig.7}
\end{picture}
\end{center}
\begin{eqnarray}
2\Sigma^{0(7)}=-\frac{M^{2}}{8L^{4/9}}
4[2(e_{u}a_{u}+e_{d}a_{d})m_{s}-
(e_{u}a_{d}+e_{d}a_{u})m_{s}+
\nonumber\\
2e_{s}(m_{u}+m_{d})a_{s}+
2(e_{u}a_{u}m_{d}+e_{d}a_{d}m_{u})-
(e_{u}m_{d}+e_{d}m_{u})a_{s}
].
\label{si7}
\end{eqnarray}
Performing changes $s\leftrightarrow d$ and $s\leftrightarrow u$ 
to obtain the auxiliary quantities we get
$$
2\tilde\Sigma^{0(7)}_{sd}=-\frac{M^{2}}{8L^{4/9}}
4[2(e_{u}a_{u}+e_{s}a_{s})m_{d}-
(e_{u}a_{s}+e_{s}a_{u})m_{d}+
$$
$$
2e_{d}(m_{u}+m_{s})a_{d}+
2(e_{u}a_{u}m_{s}+e_{s}a_{s}m_{u})-(e_{u}m_{s}+e_{s}m_{u})a_{d}],
$$
$$
2\tilde\Sigma^{0(7)}_{su}=-\frac{M^{2}}{8L^{4/9}}
4[2(e_{d}a_{d}+e_{s}a_{s})m_{u}-
(e_{d}a_{s}+e_{s}a_{d})m_{u}+
$$
$$
2e_{u}(m_{d}+m_{s})a_{u}+
2(e_{d}a_{d}m_{s}+e_{s}a_{s}m_{d})-(e_{d}m_{s}+e_{s}m_{d})a_{u}]
$$
and using Eq.(\ref{crossSi}) we obtain

$$
\Lambda^{(7)}=-\frac{M^{2}}{8L^{4/9}}
\frac{2}{3}[6(e_{u}a_{u}m_{d}+e_{d}a_{d}m_{u})+
6e_{s}a_{s}(m_{u}+m_{d})-a_{s}(e_{u}m_{d}+e_{d}m_{u})+
$$
$$
6m_{s}(e_{u}a_{u}+e_{d}a_{d})-m_{s}(e_{u}a_{d}+e_{d}a_{u})]
$$
$$
\rightarrow (\sim m_{s}, a_{u}=a_{d}=a ) -\frac{15M^{2}}{36L^{4/9}}
m_{s}(e_{u}+e_{d})a,
$$
and here it is the only discrepancy between our result
and the corresponding 
term in \cite{Pasu}, 
as they have (19/36) instead of our (15/36).The only divergence
in the 7th term seems to be due either to
some misprint or eventually to some diagram we have
not taken into account in the Fig.7. But the example of all the
other contributions including those given by the
diagrams of the Fig.9 to see later ( the contributions of 
which vanishes for all baryons but $\Lambda$ in the limit 
of zero masses of the 
light quarks) teaches us that an eventually missed diagram would
follow Eq.(\ref{crossSi}) all the same.

We have checked our result upon applying the Eq.(\ref{crossLa}) 
to the obtained expression for $\Lambda^{(7)}$ in order
to arrive at the expression for $\Sigma^{0(7)}$. The answer
have coincided with that of the Eq.(\ref{si7}).

The difference $\tilde\Sigma^{0(7)}_{su}-\tilde\Sigma^{0(7)}_{sd}$
($\sim m_{s}$) upon using Eq.(\ref{crossSiLa}) yields:
$$
\sqrt{3}(\Sigma^{0}\Lambda)^{(7)}=-\frac{M^{2}}{4L^{4/9}}
[(e_{u}a_{d}-e_{d}a_{u})m_{s}]
$$
$$
\rightarrow (\quad a_{u}=a_{d}=a) \frac{M^{2}}{L^{4/9}}
(-1/4)(e_{u}-e_{d})m_{s}a.
$$
coefficient (-1/4) to compare with (1/3) in \cite{Hwang2}.
Here again we see a discrepancy between our result and that of
\cite{Hwang2}, while as we have seen and shall
see all other terms perfectly satisfy Eq.(\ref{crossSiLa}).

The 8th term comes from the diagrams of the Fig.8:
\vskip 5mm
\begin{center}
\begin{picture}(400,100)(0,0)
\CArc(20,40)(20,90,270)
\CArc(60,40)(20,270,90)
\Line(20,60)(60,60)
\Line(20,20)(30,20)
\Line(50,20)(60,20)
\Line(0,40)(80,40)
\Photon(50,60)(50,90)3 3
\Text(40,60)[c]{$\times$}
\Text(25,70)[c]{u}
\Text(30,50)[c]{d}
\Text(50,10)[c]{s}
\CArc(220,40)(20,90,270)
\CArc(260,40)(20,270,90)
\Line(220,60)(260,60)
\Line(220,20)(230,20)
\Line(250,20)(260,20)
\Line(200,40)(280,40)
\Photon(250,60)(250,90)3 3
\Text(240,60)[c]{$\times$}
\Text(225,70)[c]{d}
\Text(230,50)[c]{u}
\Text(250,10)[c]{s}
\Text(140,0)[c]{Fig.8}
\end{picture}
\end{center}
$$
2\Sigma^{0(8)}=\frac{M^{2}}{2L^{4/9}}
[ln[\frac{M^{2}}{\Lambda^{2}}]
-1-\gamma_{EM}]2(e_{u}m_{u}+e_{d}m_{d})a_{s}
$$
where from upon using Eq.(\ref{crossSi}):
$$
\Lambda^{(8)}=\frac{M^{2}}{6L^{4/9}}[ln(\frac{M^{2}}
{\Lambda^{2}})-1-\gamma_{EM}]
[e_{u}m_{u}(2a_{d}-a_{s})+
$$
$$
e_{d}m_{d}(2a_{u}-a_{s})+2e_{s}m_{s}
(a_{u}+a_{d})]\rightarrow 
\frac{2m_{s}ae_{s}M^{2}}{3L^{4/9}}
[ln(\frac{M^{2}}{\Lambda^{2}})
-1-\gamma_{EM}]
$$
in agreement with the result of \cite{Pasu,Pasu3}.

The difference $\tilde\Sigma^{0(8)}_{su}-\tilde\Sigma^{0(8)}_{sd}$
upon using Eq.(\ref{crossSiLa}) yields:
$$
\sqrt{3}(\Sigma^{0}\Lambda)^{(8)}=
\frac{M^{2}}{2L^{4/9}}
[ln[\frac{M^{2}}{\Lambda^{2}}]
-1-\gamma_{EM}][(e_{u}m_{u}a_{d}-e_{d}m_{d}a_{u})+
e_{s}m_{s}(a_{d}-a_{u})]\Rightarrow 0.
$$

The 9th term $\Sigma^{0(9)}$ comes from the diagrams of the Fig.9:
\begin{center}
\begin{picture}(400,100)(0,0)
\CArc(20,40)(20,90,270)
\CArc(60,40)(20,270,90)
\Line(20,60)(30,60)
\Line(50,60)(60,60)
\Line(20,20)(60,20)
\Line(0,40)(80,40)
\Photon(40,60)(40,90)3 3
\Text(40,40)[c]{$\times$}
\Text(25,70)[c]{u}
\Text(30,50)[c]{d}
\Text(50,10)[c]{s}
\CArc(120,40)(20,90,270)
\CArc(160,40)(20,270,90)
\Line(120,60)(130,60)
\Line(150,60)(160,60)
\Line(120,20)(160,20)
\Line(100,40)(180,40)
\Photon(140,60)(140,90)3 3
\Text(140,40)[c]{$\times$}
\Text(125,70)[c]{d}
\Text(130,50)[c]{u}
\Text(150,10)[c]{s}
\CArc(220,40)(20,90,270)
\CArc(260,40)(20,270,90)
\Line(220,60)(230,60)
\Line(250,60)(260,60)
\Line(220,20)(260,20)
\Line(200,40)(280,40)
\Photon(240,65)(240,95)3 3
\Text(240,60)[c]{$\times$}
\Text(220,70)[c]{u}
\Text(230,50)[c]{d}
\Text(250,10)[c]{s}
\CArc(320,40)(20,90,270)
\CArc(360,40)(20,270,90)
\Line(320,60)(330,60)
\Line(350,60)(360,60)
\Line(320,20)(360,20)
\Line(300,40)(380,40)
\Photon(340,65)(340,95)3 3
\Text(340,60)[c]{$\times$}
\Text(330,70)[c]{d}
\Text(330,50)[c]{u}
\Text(320,10)[c]{s}
\Text(190,0)[c]{Fig.9}
\end{picture}
\end{center}
$$
2\Sigma^{0(9)}=-\frac{M^{4}}{4L^{28/27}}
[2(e_{u}a_{u}\chi_{u}m_{d}+e_{d}a_{d}\chi_{d}m_{u})-
2(e_{u}a_{u}\chi_{u}m_{u}+e_{d}a_{d}\chi_{d}m_{d})].
$$
This term is identically zero with 
$m_{u}=m_{d}$. So a
contribution to the $\Lambda^{(9)}$ term comes only from the
auxiliary quantities $\tilde\Sigma^{0(9)}_{sd}$ 
and $\tilde\Sigma^{0(9)}_{su}$ :
$$
2\tilde\Sigma^{0(9)}_{sd}=-\frac{M^{4}}{4L^{28/27}}
[2(e_{u}a_{u}\chi_{u}m_{s}+e_{s}a_{s}\chi_{s}m_{u})-
2(e_{u}a_{u}\chi_{u}m_{u}+e_{s}a_{s}\chi_{s}m_{s})],
$$
$$
2\tilde\Sigma^{0(9)}_{su}=-\frac{M^{4}}{4L^{28/27}}
[2(e_{d}a_{d}\chi_{d}m_{s}+e_{s}a_{s}\chi_{s}m_{d})-
2(e_{d}a_{d}\chi_{d}m_{d}+e_{s}a_{s}\chi_{s}m_{s})].
$$
As a result we get from Eq.(\ref{crossSi}):
$$
\Lambda^{(9)}=
-\frac{M^{4}}{12L^{28/27}}
\{2 [e_{u}a_{u}\chi_{u}+e_{d}a_{d}\chi_{d}-
2e_{s}a_{s}\chi_{s}]m_{s}-
(e_{u}a_{u}\chi_{u}m_{d}+e_{d}a_{d}\chi_{d}m_{u})-
$$
$$
(e_{u}a_{u}\chi_{u}m_{u}+e_{d}a_{d}\chi_{d}m_{d})+
2e_{s}a_{s}\chi_{s}(m_{u}+m_{d})\}
$$
and for $m_{u}=m_{d}=0$,$a_{u}=a_{d}=a$,
$\chi_{u}=\chi_{d}=\chi$, $\chi_{s}a_{s}/\chi a=\phi$
$$
\rightarrow
-\frac{M^4}{6L^{28/27}}(e_{u}+e_{d}-2e_{s}\phi)
m_{s}a\chi
$$
in accord with the Eq.(4) in \cite{Pasu2}.

The difference $\tilde\Sigma^{0(9)}_{su}-\tilde\Sigma^{0(9)}_{sd}$
upon using Eq.(\ref{crossSiLa}) yields:
$$
\sqrt{3}(\Sigma^{0}\Lambda)^{(9)}=\frac{M^{4}}{4L^{28/27}}
[m_{s}a\chi(e_{u}-e_{d})+e_{s}a_{s}\chi_{s}(m_{u}-m_{d})-
(e_{u}m_{u}-e_{d}m_{d})a\chi]
$$
and for $m_{u}=m_{d}=0$, 
$$
\rightarrow-\frac{M^{2}}{4L^{28/27}}m_{s}a\chi(e_{u}-e_{d}),
$$
which agrees with the corresponding term in \cite{Hwang2}.

The 10th term $\Sigma^{0(10)}$ 
comes from the diagrams of the Fig.10:
\vskip 5mm
\begin{center}
\begin{picture}(400,100)(0,0)
\CArc(20,40)(20,90,270)
\CArc(60,40)(20,270,90)
\Line(20,60)(30,60)
\Line(50,60)(60,60)
\Line(20,20)(60,20)
\Line(0,40)(80,40)
\DashLine(40,20)(40,0)2
\Photon(40,60)(40,90)3 3
\Text(40,40)[c]{$\times$}
\Text(20,70)[c]{u,d}
\Text(25,50)[c]{d,u}
\Text(60,10)[c]{s}
\CArc(120,40)(20,90,270)
\CArc(160,40)(20,270,90)
\Line(120,60)(160,60)
\Line(120,20)(160,20)
\Line(100,40)(130,40)
\Line(150,40)(180,40)
\Photon(140,40)(160,70)3 3
\DashCArc(130,40)(10,180,0)2
\Text(140,60)[c]{$\times$}
\Text(120,70)[c]{u,d}
\Text(120,50)[c]{d,u}
\Text(170,10)[c]{s}
\CArc(220,40)(20,90,270)
\CArc(260,40)(20,270,90)
\Line(220,60)(230,60)
\Line(250,60)(260,60)
\Line(220,20)(260,20)
\Line(200,40)(280,40)
\Photon(240,60)(240,90)3 3
\DashLine(220,40)(240,55)2
\Text(250,40)[c]{$\times$}
\Text(220,70)[c]{u,d}
\Text(220,50)[c]{d,u}
\Text(260,10)[c]{s}
\Text(140,0)[c]{Fig.10}
\end{picture}
\end{center}
The last diagram contributes only to the $\kappa$ term and has an
infrared divergence regularized by a cutoff $\Lambda$. Finally,
$$
2\Sigma^{0(10)}=
[\frac{1}{6}(2\kappa_{u}-\xi_{u}))M^{2}-\frac{M^{2}}{2}\kappa_{u}
[ln[\frac{M^{2}}{\Lambda^{2}}]
-1-\gamma_{EM}]]
e_{u}a_{u}m_{d}+
$$
$$
[\frac{1}{6}(2\kappa_{d}-\xi_{d}))M^{2}-\frac{M^{2}}{2}\kappa_{d}
[ln[\frac{M^{2}}{\Lambda^{2}}]
-1-\gamma_{EM}]]
e_{d}a_{d}m_{u}.
$$
Performing changes $(d\leftrightarrow s )$ and 
$(u\leftrightarrow s )$ to construct quantities
$\tilde\Sigma^{0(10)}_{sd}$ and $\tilde\Sigma^{0(10)}_{su}$ 
and putting them into Eq.(\ref{crossSi})
we obtain for the contribution $\Lambda^{(10)}$ 
$$
\Lambda^{(10)}=
M^{2}[\frac{1}{36}(2\kappa_{u}-\xi_{u}))-\frac{1}{12}\kappa_{u}
[ln(\frac{M^{2}}{\Lambda^{2}})
-1-\gamma_{EM}]]
e_{u}a_{u} (2m_{s}-m_{d})+
$$
$$
M^{2}[\frac{1}{36}(2\kappa_{d}-\xi_{d}))-\frac{1}{12}\kappa_{d}
[ln(\frac{M^{2}}{\Lambda^{2}})
-1-\gamma_{EM}]]
e_{d}a_{d} (2m_{s}-m_{u})+
$$
$$
2M^{2}[\frac{1}{36}(2\kappa_{s}-\xi_{s}))-\frac{1}{12}\kappa_{s}
[ln(\frac{M^{2}}{\Lambda^{2}})
-1-\gamma_{EM}]]
e_{s}a_{s} (m_{u}+m_{d}).
$$
or (maintaining only terms $\sim m_{s}$
and putting $a_{u}=a_{d}=a$,
$\kappa_{u}=\kappa_{d}=\kappa $, $\xi_{u}=\xi_{d}=\xi$):
$$
\Lambda^{(10)}=(e_{u}+e_{d})
[\frac{1}{18}(2\kappa-\xi))M^{2}-\frac{M^{2}}{6}\kappa
[ln(\frac{M^{2}}{\Lambda^{2}})
-1-\gamma_{EM}]]a m_{s} 
$$
again in full agreement with the results of \cite{Pasu},\cite{Pasu3}.

The corresponding term for $\Sigma^{0}\Lambda$ 
transition magnetic moment can be readily
obtained through Eq.(\ref{crossSiLa}):
$$
\sqrt{3}(\Sigma^{0}\Lambda)^{(10)}=
[\frac{1}{12}(2\kappa-\xi)M^{2}-\frac{M^{2}}{4}\kappa
[ln(\frac{M^{2}}{\Lambda^{2}})
-1-\gamma_{EM}]][m_{s}a(e_{u}-e_{d})+
$$
$$
e_{s}a_{s}\chi_{s}(m_{u}-m_{d})]
\rightarrow 
(e_{u}-e_{d})[\frac{1}{12}(2\kappa-\xi)M^{2}-\frac{M^{2}}{4}\kappa
[ln(\frac{M^{2}}{\Lambda^{2}})
-1-\gamma_{EM}]]m_{s}a.
$$
Combining all the terms $\Lambda^{(k)}, k=1,...,10,$ 
we arrive at the QCD Borel sum rule
for the $\Lambda$ magnetic moment:
$$
\frac{M^{6}}{12L^{4/9}}(e_{u}+e_{d}+4e_{s})+
\frac{M^{2}b}{144L^{4/9}}(4e_{u}+4e_{d}+7e_{s})
$$
$$
-\frac{L^{4/9}}{108M^{2}}
[2(7e_{u}+7e_{d}+e_{s})+8f(2e_{u}+2e_{d}+e_{s})]a^{2}-
$$
$$
\frac{M^{2}b}{192L^{4/9}}[ln(\frac{M^{2}}{\Lambda^{2}})
-1-\gamma_{EM}]
(5e_{u}+5e_{d}+2e_{s})-
$$
$$
\frac{M^{2}b}{108L^{4/9}}[ln(\frac{M^{2}}{\Lambda^{2}})
-\gamma_{EM}-\frac{M^{2}}{2\Lambda^{2}}]
(e_{u}+e_{d}+4e_{s})+
$$
$$
[\frac{-\chi a^{2}}{18 L^{4/27}}
(M^{2}-\frac{m_{(s)0}^{2}}{8L^{4/9}})+\frac{L^{4/9}}{108}
(2\kappa-\xi )a^{2}][(e_{u}+e_{d})(1+2f)+4e_{s}\phi]
$$
$$
-\frac{15M^{2}}{36L^{4/9}}
m_{s}(e_{u}+e_{d})a+
\frac{2m_{s}ae_{s}M^{2}}{3L^{4/9}}
[ln(\frac{M^{2}}{\Lambda^{2}})
-1-\gamma_{EM}]
$$
$$
-\frac{M^2}{6L^{28/27}}(e_{u}+e_{d}-2e_{s}\phi)m_{s}a\chi+
$$
$$
(e_{u}+e_{d})
[\frac{1}{18}(2\kappa-\xi))M^{2}-\frac{M^{2}}{6}\kappa
[ln(\frac{M^{2}}{\Lambda^{2}})
-1-\gamma_{EM}]]a m_{s} 
$$
$$
=\beta_{\Lambda}^{2}
\mu_{\Lambda}e^{-M_{\Lambda}^{2} /M^{2} }
(1+A_{\Lambda}M^{2} )+e.s.c.
$$
in agreement with \cite{Pasu}
but the coefficient (15/36) instead of (19/36) in the 7th term. 

Combining all the terms $(\Sigma^{0}\Lambda)^{k}, k=1,...10,$ 
we arrive at the QCD Borel sum rule
for the $\Sigma^{0}\Lambda$ transition magnetic moment:
\begin{eqnarray}
(e_{u}-e_{d})[\frac{M^{6}}{4L^{4/9}}+
\frac{L^{4/9}}{9M^{2}}a_{s}a+
\frac{M^{2}b}{24L^{4/9}}
\nonumber\\
-\frac{M^{2}b}{144L^{4/9}}
[ln(\frac{M^{2}}{\Lambda^{2}})
-1-\gamma_{EM}]+
4[ln(\frac{M^{2}}{\Lambda^{2}})
-\gamma_{EM}-\frac{M^{2}}{2\Lambda^{2}}]+[
\nonumber\\
 \frac{-\chi }{6L^{4/27}}
(M^{2}-\frac{m_{(s)0}^{2}}{8L^{4/9}})+\frac{L^{4/9}}{36}
(2\kappa-\xi )]a a_{s}+
\nonumber\\
\frac{M^{2}}{L^{4/9}}\frac{1}{4}
m_{s}a-
\frac{M^{4}}{4L^{28/27}}m_{s}a\chi+
[\frac{1}{18}(2\kappa-\xi)M^{2}-
\nonumber\\
\frac{M^{2}}{6}\kappa
[ln(\frac{M^{2}}{\Lambda^{2}})
-1-\gamma_{EM}]]m_{s}a
=\beta_{\Sigma}\beta_{\Lambda}\sqrt{3}
\mu_{\Sigma^{0}\Lambda}e^{-\bar m^{2} /M^{2} }
(1+A_{\Sigma^{0}\Lambda}M^{2} )+e.s.c.
\end{eqnarray}

\section{Conclusion}

We have shown on the example of 
QCD Borel sum rules for magnetic moments 
of the $\Sigma$ and $\Lambda$ hyperons 
in the version proposed in \cite{Pasu} that starting
from the sum rule for the $\Sigma$ 
hyperon it is straightforward to obtain the corresponding
sum rule for the $\Lambda$ hyperon
upon using intercrossed relation Eq.(\ref{crossSi}) as well 
as to construct a corresponding sum rule for
the $\Lambda-\Sigma^{0}$ transition magnetic moment.

We have also checked
without writing it down that starting from the obtained
QCD sum rule for the $\Lambda$ magnetic moment and
applying Eq.(\ref{crossLa}) we return to the initial sum rule
for the $\Sigma^{0}$ magnetic moment.

We have checked these formulae for the QCD Light-Cone sum rules
written by one of us with coauthors \cite{Altug} and
seen that the Eqs.((\ref{crossSi},\ref{crossLa}) are
satisfied exactly. The results will be published elsewhere.
But we think that it is more convincing and instructive to take 
already published results in order to prove our formulae.
Earlier similar result has been proved for the QCD mass sum 
rules \cite{OYZ} upon relating $\Sigma^{0}$ and $\Lambda$
mass sum rules where the agreement with the sum rules of 
\cite{Hwang} has been found to be perfect.

As for magnetic moment sum rules
we have shown explicitly that every group
of diagrams for $\Sigma^{0}$ generates according to
Eqs.(\ref{crossSi},\ref{crossSiLa}) analogous
groups of diagrams for $\Lambda$ and $\Sigma^{0}\Lambda$
in practically full agreement with the results of
\cite{Pasu2} and \cite{Hwang2}.
( The only our divergence with \cite{Pasu2}
in the 7th term seems to be due either to
some misprint or eventually to some diagram we have
not taken into account.)

The proposed relations Eqs.(\ref{crossSi}-\ref{crossLaSi})
can be used not only to obtain many properties
of the the $\Sigma$-like baryons
from those of $\Lambda$-like ones {\it et vice versa}
but also to check mutually many-terms relations for
the $\Sigma$-like and $\Lambda$-like baryons.

\vskip 5mm
{\large{\bf Acknowledgments}}
\vskip 5mm

We thank B.L.Ioffe for fruitful discussion.
We are grateful to T.Aliev, F.Hussein, G.Thompson
for their interest to the work. 

One of us (V.Z.) is grateful to Prof.S.Randjbar-Daemi
for the hospitality extended to him at HE section of ICTP 
(Trieste, Italy). The financial support of the ICTP in 
the initial stage of the work is gratefully acknowledged.

%
%


\newpage

\begin{thebibliography}{99}
\bibitem{Hwang}W-Y.P.Hwang and K.-C.Yang, Phys.Rev.D{\bf  49},460 (1994).
\bibitem{Hwang1} K.-C.Yang, W-Y.Hwang, E.M.Henley, and 
L.S.Kisslinger, Phys.Rev.D{\bf  47},3001 (1993).
\bibitem{Hwang2} Shi-lin Zhu, W-Y.P.Hwang and Ze-sen Yang, 
Phys.Rev.D{\bf 57}, 1527 (1998).
\bibitem{Altug1} T.M.Aliev, A.\"{O}zpineci, M.Savci, 
Phys.Rev.D{\bf 62}, 053012 (2000).
\bibitem{Altug} T.M.Aliev, A.\"{O}zpineci, M.Savci, 
Nucl. Phys. {\bf A678}, 443 (2000).
\bibitem{Ioffe} V.M.Belyaev, B.L.Ioffe, JETP {\bf 56}, 493 (1982);
B.L.Ioffe, Smilga, Nucl.Phys.  {\bf B232}, 109 (1984);
I. I. Balitsky and A. V. Yung,
Phys. Lett. {\bf B129}, 328 (1983).
\bibitem{Pasu} Ch.B.Chiu, J.Pasupathy, S.L.Wilson,
Phys.Rev.D{\bf 33},1961 (1986).
\bibitem{Pasu2} Ch.B.Chiu, J.Pasupathy, S.L.Wilson,
Phys.Rev.D{\bf 36},1442 (1987).
\bibitem{Pasu3} Ch.B.Chiu, S.L.Wilson, J.Pasupathy, J.P.Singh,
Phys.Rev.D{\bf 36},1553 (1987).
\bibitem{OYZ} A.\"{O}zpineci, S.B.Yakovlev, V.S.Zamiralov, 
NPI MSU Preprint 
2003-18/731, Skobeltsyn Institute of Nuclear Physics,
Moscow State University, Moscow, Russia.
\end{thebibliography}
\end{document}